\documentclass[12pt]{article}
\usepackage[reqno]{amsmath}
\usepackage{bbm}
\usepackage{epsfig}
\usepackage{array}
\usepackage{float}
\usepackage{dsfont}
\usepackage{amstext}
\usepackage{amsfonts}

\usepackage{cite}

\usepackage{a4}

\usepackage{a4wide}
\def\be{\begin{equation}}
\def\ee{\end{equation}}
\def\gs{\mathrel{
   \rlap{\raise 0.511ex \hbox{$>$}}{\lower 0.511ex \hbox{$\sim$}}}}
\def\ls{\mathrel{
   \rlap{\raise 0.511ex \hbox{$<$}}{\lower 0.511ex \hbox{$\sim$}}}}

\newcommand{\onbb}{neutrinoless double beta decay}
\newcommand{\ba}{\begin{array}{c}}
\newcommand{\baz}{\begin{array}{cc}}
\newcommand{\bad}{\begin{array}{ccc}}
\newcommand{\bav}{\begin{array}{cccc}}
\newcommand{\bea}{\begin{equation} \begin{array}{c}}
\newcommand{\eea}{ \end{array} \end{equation}}
\newcommand{\ea}{\end{array}}
\newcommand{\D}{\displaystyle}

\newcommand{\dma}{\mbox{$\Delta m^2_{\rm A}$}}

\begin{document}

\title{
\hfill {\small arXiv: 0810.5239 [hep-ph]} 
\vskip 0.4cm
\bf 
Unified Parametrization for Quark and Lepton Mixing Angles
}
\author{
Werner Rodejohann\thanks{email: \tt werner.rodejohann@mpi-hd.mpg.de} 
\\\\
{\normalsize \it Max--Planck--Institut f\"ur Kernphysik,}\\
{\normalsize \it  Postfach 103980, D--69029 Heidelberg, Germany}
}
\date{}
\maketitle
\thispagestyle{empty}
\begin{abstract}
\noindent  
We propose a new parametrization for the quark and lepton mixing
matrices: the two 12-mixing angles 
(the Cabibbo angle and the angle responsible for solar 
neutrino oscillations) are at zeroth order $\pi/12$ and $\pi/5$, 
respectively. The resulting 12-elements in the CKM 
and PMNS matrices, $V_{us}$ and $U_{e2}$,  
are in this order irrational but simple algebraic numbers. 
We note that the cosine of $\pi/5$ is the golden ratio divided by two. 
The difference between $\pi/5$ and the 
observed best-fit value of solar neutrino mixing is of the same order 
as the difference between 
the observed value and the one for tri-bimaximal mixing. In order to 
reproduce the central values of current fits,  
corrections to the zeroth order expressions are necessary. 
They are small and of the same order and sign for quarks and leptons. 
We parametrize the perturbations to the CKM and
PMNS matrices in a ``triminimal'' way, i.e., with three small
rotations in an order corresponding to the order of the 
rotations in the PDG-description of
mixing matrices.

\end{abstract}

\newpage

Quark and lepton mixing can successfully be described in the Standard 
Model of elementary particle physics. For each sector a unitary 
mixing matrix connects mass and flavor states, 
and can be parametrized as \cite{PDG}
\bea  \label{eq:PDG}
V, \, U  
 = R_{23}(\theta_{23}^{q,\ell}) 
\, R_{13}^{q,\ell}(\theta_{13}^{q,\ell} ; \delta^{q,\ell}) \, 
R_{12}(\theta_{12}^{q,\ell}) \\ 
 =
\left( \bad 
c_{12}^{q,\ell}  \, c_{13}^{q,\ell} 
& s_{12}^{q,\ell} \, c_{13}^{q,\ell} 
& s_{13}^{q,\ell} \, e^{-i \delta^{q,\ell}}  \\ 
-s_{12}^{q,\ell} \, c_{23}^{q,\ell} 
- c_{12}^{q,\ell} \, s_{23}^{q,\ell} \, 
s_{13}^{q,\ell}  \, e^{i \delta^{q,\ell}} 
& c_{12}^{q,\ell} \, c_{23}^{q,\ell} - 
s_{12}^{q,\ell} \, s_{23}^{q,\ell} \, s_{13}^{q,\ell} 
\, e^{i \delta^{q,\ell}} 
& s_{23}^{q,\ell}  \, c_{13}^{q,\ell}  \\ 
s_{12}^{q,\ell}   \, s_{23}^{q,\ell} - c_{12}^{q,\ell} 
\, c_{23}^{q,\ell}  \, s_{13}^{q,\ell} \, e^{i \delta^{q,\ell}} & 
- c_{12}^{q,\ell} \, s_{23}^{q,\ell} 
- s_{12}^{q,\ell} \, c_{23}^{q,\ell} \, 
s_{13}^{q,\ell} \, e^{i \delta^{q,\ell}} 
& c_{23}^{q,\ell}  \, c_{13}^{q,\ell}  
\ea   
\right) P^{q,\ell}\,.
\eea
Here $V$ is the Cabibbo-Kobayashi-Maskawa matrix (CKM) 
matrix containing the 
mixing angles $\theta_{ij}^q$ and $U$ is the 
Pontecorvo-Maki-Nakagawa-Sakata (PMNS) matrix containing the 
mixing angles $\theta_{ij}^\ell$. As usual, $c_{ij}^{q,\ell} = 
\cos \theta_{ij}^{q,\ell}$ and 
$s_{ij}^{q,\ell} = \sin \theta_{ij}^{q,\ell}$. There is also a diagonal
phase matrix, which is trivial in the quark sector 
$(P^q = \mathbbm{1})$, and contains the Majorana phases in 
the lepton sector: $P^\ell = {\rm diag}(1,e^{i\alpha},e^{i\beta})$. 
As indicated, the above mixing matrices are products of 
rotations, e.g. 
\[ 
R_{23}(\theta) = 
\left( 
\bad
1 & 0 & 0 \\
0 & \cos \theta & \sin \theta \\
0 & -\sin \theta & \cos \theta 
\ea
\right)\,,~~ R_{13}(\theta ; \delta)  = 
\left( 
\bad
\cos \theta & 0 & \sin \theta \, e^{-i\delta}  \\
0 & 1 & 0 \\ 
-\sin \theta \, e^{i\delta} & 0 & \cos \theta \\ 
\ea
\right) .
\]

Impressive progress has been achieved in determining the parameters 
of $U$ and $V$. 
The current knowledge for the quark sector 
can be summarized at 1$\sigma$ as \cite{PDG}
\be\label{eq:dataq}
\bad
\sin \theta_{12}^q &=& 0.2257 \pm 0.0010 \,,\\ [0.2cm] 
\sin \theta_{23}^q &=& 0.0415 ^{+0.0010}_{-0.0011} \,,\\[0.2cm] 
\sin \theta_{13}^q &=& 0.00359 \pm 0.00016 \,.
\ea
\ee
The CP phase lies in the range (see also e.g.~\cite{CKMfitter}) 
$\delta^q = \left(68.85^{+3.04}_{-5.42}\right)^\circ$. 
Regarding the leptons, we have at 1, 2 and $3\sigma$ \cite{bari}
\be \label{eq:datanu}
\bad
\sin \theta_{12}^\ell &=& 
0.559^{+0.017, \, 0.035, \, 0.054}_{-0.016, \, 0.031, \, 0.046} 
\,,\\[0.2cm] 
\sin \theta_{23}^\ell &=& 
0.683^{+0.052, \, 0.093, \, 0.120}_{-0.044, \, 0.078, \, 0.107} 
\,,\\[0.2cm] 
\sin \theta_{13}^\ell &=& 
0.126^{+0.035, \, 0.063, \, 0.088}_{-0.049, \, 0.126, \, 0.126} \,.
\ea
\ee
The hint for non-zero $\theta_{13}^\ell$ \cite{bari} 
(see also \cite{others}) is rather weak, surviving only $1.6 \sigma$. 
There is no information on leptonic CP violation.\\ 

The precision era which also neutrino physics has entered recently 
allows in particular 
to parametrize the PMNS matrix with reasonable accuracy.  
Trying to parametrize the mixing matrices 
\cite{wolf,KM,Xing,Zee,WR,chin0,ER,Trim2,SFK,PRW,web} 
might be helpful phenomenologically as well as could hint 
towards a structure underlying the observed mixing patterns. 
There are three desiderata for convenient parametrizations:
\begin{itemize}
\item[a)] fast convergence, i.e., the zeroth order 
expression should be close 
to the observed values; 
\item[b)] at zeroth order the mixing matrix entries should be simple 
numbers; 
\item[c)] similar parametrizations 
should be used for both quarks and leptons. 
\end{itemize} 
The latter implies usually that very different zeroth order 
forms of $U$ and $V$ are needed. For
instance, one could use the unit matrix for the quarks and
tri-bimaximal \cite{tri} mixing (or bimaximal \cite{bi}) 
for the leptons. In this letter we will propose to use in the zeroth 
order matrices angles which are fractions of $\pi$ with an integer
number. This is of course common practice for atmospheric mixing, 
$(\theta_{23}^\ell)^0 = \pi/4$, and will be used here for the 12-rotations
of the quarks ($(\theta_{12}^q)^0 = \pi/12$) and leptons 
($(\theta_{12}^\ell)^0 = \pi/5$) as well. 
The value  $(\theta_{12}^\ell)^0 = \pi/5$ is 
in fact within the allowed 
$2\sigma$ range. 
Our alternative Ansatz brings the zeroth order forms of $U$ and $V$ on 
somewhat equal footing and can be used for the quark and 
lepton sectors simultaneously. 
Of course, the angles themselves are 
not physical and our phenomenological Ansatz relies on the 
particular parametrization of Eq.~(\ref{eq:PDG}). 
Nevertheless, the sines and cosines of the angles are almost the 
mixing matrix elements, which are indeed physical quantities\footnote{ 
Ref.~\cite{Trim2} has recently proposed to use the value 
$\sin (\theta_{12}^q)^0 = \frac{\sqrt{2} - 1}{\sqrt{6}}$, 
or $(\theta_{12}^q)^0 \simeq \pi/18.49$, as the zeroth 
order expression for the Cabibbo angle.}. 
With the angles chosen here, the sines and cosines turn out to be 
simple and concise irrational numbers. This is reminiscent of 
bimaximal or tri-bimaximal mixing. In addition, comparing our
proposal with current best-fit values reveals that the required perturbation 
parameters are small and for the lepton sector of the same order as
the required perturbation parameters for deviations from 
tri-bimaximal mixing. 
They are furthermore of the same order and sign for both quarks
and leptons, at least for the 12-sector. We have thus met all three 
desiderata given above.

Having a zeroth order form of the mixing matrix implies that in
general (ignoring the Majorana phases of the lepton sector) four small
parameters $\epsilon_{ij}^{q,\ell}$ have to be introduced, 
corresponding to the four parameters describing all mixing 
phenomena. If the four parameters are 
introduced in terms of (small angle) rotations, it is most
straightforward to choose the order of rotations such that 
each small parameter $\epsilon_{ij}^{q,\ell}$ is responsible 
for the deviation of (and only of) $\theta_{ij}^{q,\ell}$ from 
its initial value. 
This ``triminimal'' parametrization \cite{PRW} has been applied for
the tri-bimaximal \cite{tri} mixing scheme (TBM). 
The latter corresponds to 
$(\theta_{23}^\ell)^0 = \pi/4$, $(\theta_{13}^\ell)^0 =
0$ and $(\theta_{12}^\ell)^0 = \theta_{\rm TBM}$, or 
$U^0 = R_{23}(\pi/4) \, R_{12}(\theta_{\rm TBM})$, where 
$\sin^2 \theta_{\rm TBM} = \frac 13$. One then parametrizes the PMNS
matrix triminimally as \cite{PRW}
\be \label{eq:PRW}
U = R_{23}(\pi/4) \, R_{23} (\epsilon_{23}^\ell) \, 
R_{13} (\epsilon_{13}^\ell ; \delta^\ell) \, R_{12}
(\epsilon_{12}^\ell) \, R_{12} (\theta_{\rm TBM})\,.
\ee
The order of small rotations in between the rotations with the 
large angles $\pi/4$ and $\theta_{\rm TBM}$ 
is the same as in the PDG-description of 
a mixing matrix. It is then easy to see that each 
$\epsilon_{ij}^\ell$ describes the deviation of (and only of) 
$\theta_{ij}^\ell$ from $(\theta_{ij}^\ell)^0$. Moreover, 
the introduced CP phase appears exactly where it appears in 
Eq.~(\ref{eq:PDG}). Note that a 
triminimal parametrization is manifestly unitary. 
If it turns out that one of the deviations from 
tri-bimaximal mixing is sizable, this parametrization 
can treat that case more accurately.  Regarding the quark sector, the
hierarchy in the CKM angles implies to start with only a non-zero
12-rotation and then introduce from the left in a triminimal way three 
small rotations in the order of the PDG parametrization. This has 
recently been proposed in Ref.~\cite{Trim2}.\\

Let us start by considering the quark sector. The goal is to find 
a simple initial mixing angle $(\theta_{12}^q)^0$, being a fraction of
$\pi$ with an integer number, and which could be used as starting point for 
an expansion. In addition it should yield a simple number for 
the sine and cosine. This leads to the 
choice\footnote{
Actually, a fraction of $\pi$ which is slightly 
closer to the measured value 
of the Cabibbo angle is $(\theta_{12}^q)^0 = \frac{\pi}{14} \simeq 0.2244$. 
We note that $V_{us} = \sin \frac{\pi}{14}$ has 
been obtained from a flavor symmetry in Refs.~\cite{MPIK}. It was shown 
that if a dihedral group $D_n$ (generated by the generators $A$ and
$B$) is broken by Higgs vevs to different $Z_2$ subgroups 
($B \,A^{m_u}$ in the up-quark sector, 
$B \,A^{m_d}$ for the down-quarks), then 
$|V_{us}| = |\cos \pi (m_u - m_d) \, j/n|$ is obtained. 
Here $j$ is the index of the representation under which the 
quarks transform. In case $j = 1$, $m_u = 3$, $m_d = 0$ and $n=7$ 
it follows that 
$|V_{us}| = \cos \frac{3\pi}{7} = \sin
\frac{\pi}{14}$. Similar considerations can be performed for other
mixing angles. Anyway, here we restrict ourselves to pure phenomenology.}
\be \label{eq:12q}
(\theta_{12}^q)^0 = \frac{\pi}{12} \Rightarrow \sin (\theta_{12}^q)^0 
= \frac{\sqrt{3} - 1}{2\sqrt{2}} = 0.2588\,,
\ee
such that at zeroth order the CKM matrix is 
\be
V^0 = 
\left( 
\bad
\frac{\sqrt{3} + 1}{2\sqrt{2}} & \frac{\sqrt{3} - 1}{2\sqrt{2}} & 0 \\
\frac{1 - \sqrt{3}}{2\sqrt{2}} & \frac{\sqrt{3} + 1}{2\sqrt{2}} & 0 \\
0 & 0 & 1 
\ea
\right) = 
\left( 
\bad
0.9659 & 0.2588 & 0 \\
-0.2588 & 0.9659 & 0 \\
0 & 0 & 1
\ea
\right) .
\ee
In the spirit of triminimality \cite{PRW}, we can then describe the 
small deviations of this matrix with 
\be
V = R_{23} (\epsilon_{23}^q) \, R_{13}
(\epsilon_{13}^q; \delta^q) \, R_{12} (\epsilon_{12}^q) \, 
V^0 \,.
\ee
As mentioned above, a triminimal parametrization has the 
obvious advantage that 
each small parameter $\epsilon_{ij}^q$ is responsible for the deviation 
of $\theta_{ij}^q$ from its initial value. In this case $\epsilon_{23}^q$ 
is $\theta_{23}^q$, and $V_{ub} = \sin \epsilon_{13}^q \, 
e^{-i \delta^q}$. The allowed ranges of $\epsilon_{13, 23}^q$ 
are nothing but the allowed ranges of the parameters 
$\theta_{13, 23}^q$ given in Eq.~(\ref{eq:dataq}). 
In addition, $\delta^q $ is directly
interpretable as the CP phase in the usual PDG parametrization. 
Expanding the CKM matrix to 
first order gives 
\bea \D 
V \simeq V^0 + \epsilon_{12}^q 
\left(
\bad 
\frac{1 - \sqrt{3}}{2\sqrt{2}} & 
\frac{1 + \sqrt{3}}{2\sqrt{2}} & 0 \\
-\frac{1 + \sqrt{3}}{2\sqrt{2}} & 
\frac{1 - \sqrt{3}}{2\sqrt{2}} & 0 \\ 
0 & 0 & 0
\ea
\right) + 
\epsilon_{13}^q
\left(
\bad 
0 & 0 & e^{-i\delta^q} \\ 
0 & 0 & 0 \\ 
-\frac{1 + \sqrt{3}}{2\sqrt{2}} \, e^{i\delta^q} & 
\frac{1 - \sqrt{3}}{2\sqrt{2}} \, e^{i\delta^q} & 0
\ea
\right) \\ \D 
+
\epsilon_{23}^q 
\left(
\bad
0 & 0 & 0 \\
0 & 0 & 1 \\ 
-\frac{1 - \sqrt{3}}{2\sqrt{2}} & 
-\frac{1 + \sqrt{3}}{2\sqrt{2}} & 0 
\ea
\right) .
\eea
The sine of the 12-mixing angle is given by 
\be
\sin \theta_{12}^q = \frac 12 \, 
\sqrt{2 - \sqrt{3} \, \cos 2 \epsilon_{12}^q + \sin 2 \epsilon_{12}^q}
\simeq \frac{\sqrt{3} - 1}{2\sqrt{2}}  
\left(
1 + (2 + \sqrt{3}) \, \epsilon_{12}^q
\right)\,.
\ee
Note that the last expression is equivalent to 
$\sin \theta_{12}^q \simeq \sin (\theta_{12}^q)^0 + \epsilon_{12}^q \,
\cos (\theta_{12}^q)^0$. 
Numerically we have $\sin \theta_{12}^q \simeq 0.2588 + 0.9659 \, 
\epsilon_{12}^q$, so that $\epsilon_{12}^q$ can be almost directly 
identified with the deviation of the sine of Cabibbo angle from 
$\frac{\sqrt{3} - 1}{2\sqrt{2}}$. In order to bring 
$\sin \theta_{12}^q$ in the observed range given in Eq.~(\ref{eq:dataq}) 
one requires 
\be \label{eq:eps12q}
\epsilon_{12}^q = -0.0341 \pm 0.0010\,.
\ee
There is a hierarchy implied for the small parameters, namely 
$(\epsilon_{12}^q)^2 \sim (\epsilon_{23}^q)^2 \sim \epsilon_{13}^q$. 
The Jarlskog invariant $J_{\rm CP}^q
= {\rm Im}\left( V_{us} \, V_{cs}^\ast 
\, V_{ub}^\ast \, V_{cb}\right)$ is 
\be
J_{\rm CP}^q = \frac {1}{16} \, 
\cos \epsilon_{13}^q \,\sin 2\epsilon_{13}^q \, \sin 2 \epsilon_{23}^q 
\, ( \cos 2 \epsilon_{12}^q + \sqrt{3} \sin 2 \epsilon_{12}^q ) \, 
\sin \delta^q\,.
\ee
In the limit of $\delta^q = \pi/3$ and $\theta_{12}^q = \pi/12$ 
we have $J_{\rm CP}^q = \frac{\sqrt{3}}{32} \, \sin 2 \theta_{23}^q 
\, \sin 2 \theta_{13}^q \, \cos \theta_{13}^q 
\simeq  \frac{\sqrt{3}}{8} \, \theta_{23}^q 
\, \theta_{13}^q$. 

In principle one could also start with non-zero zeroth order expressions for 
$\sin \theta_{23, 13}^q$ in which the angles are also written as $\pi/n$, 
for instance $ \theta_{23}^q = \frac{\pi}{76} \simeq 0.0413$ 
and $\theta_{13}^q = \frac{\pi}{875} \simeq 0.00359$.  
Since $\theta_{23, 13}^q$ are very small, 
this is not necessary. Note however that $\delta^q = \pi/3$ 
is a good approximation.\\

Turning to the lepton sector, it is trivial to note that 
atmospheric neutrino mixing can be described with 
$(\theta_{23}^\ell)^0 = \pi/4$ and that $(\theta_{13}^\ell)^0 = 0$. 
In analogy to the 
discussion for the quark sector, we propose to use as 
the zeroth order expression for the (solar) 12-rotation 
\be
(\theta_{12}^\ell)^0 = \frac{\pi}{5} \Rightarrow 
\sin^2 (\theta_{12}^\ell)^0 = \frac{5 - \sqrt{5}}{8} 
\simeq 0.345\,.
\ee
As can be seen from Eq.~(\ref{eq:datanu}), 
this value is within the $2\sigma$ range of 
the oscillation parameters. The corresponding 12-rotation is 
\be
R_{12}(\pi/5) = 
\frac 14 
\left( 
\bad
1 + \sqrt{5} & \sqrt{2}\sqrt{5 - \sqrt{5}} & 0 \\
-\sqrt{2}\sqrt{5 - \sqrt{5}} & 1 + \sqrt{5} & 0 \\
0 & 0 & 4 
\ea
\right) = 
\left( 
\bad
0.809 & 0.588 & 0 \\
-0.588 & 0.809 & 0 \\
0 & 0 & 1
\ea
\right) .
\ee
Note that $\cos \frac \pi5 = \varphi/2$, 
where $\varphi = \frac 12 \, (1 + \sqrt{5})$ is the golden 
ratio\footnote{Ref.~\cite{verrueckt} has proposed to 
identify the cotangent of $\theta_{12}^\ell$ 
with the golden ratio. That this value is close to 
the best-fit point has been also observed in 
Ref.~\cite{other_golden}. We note 
that our choice of $\cos \theta_{12}^\ell = \varphi/2$ is 
closer to the best-fit point.}. 
Consequently, $\sin^2 \frac \pi5 = \frac 14 \, (3 - \varphi)$,  
$\sin^2 2\frac{\pi}{5} = \frac{\sqrt{5}}{4} \, \varphi$, etc. 
The zeroth order PMNS matrix is (omitting the Majorana phases) 
\be
U^0 = 
\begin{pmatrix}
\frac{1 + \sqrt{5}}{4} & \frac{\sqrt{5-\sqrt{5}}}{2\sqrt{2}}  & 0 \\
-\frac{\sqrt{5- \sqrt{5}}}{4} & 
\frac{1 + \sqrt{5}}{4\sqrt{2}}  & -\frac{1}{\sqrt{2}} \\
-\frac{\sqrt{5-\sqrt{5}}}{4} & 
\frac{1 + \sqrt{5}}{4\sqrt{2}} & \frac{1}{\sqrt{2}}
\end{pmatrix}
= 
\begin{pmatrix}
0.809 & 0.588  & 0 \\
-0.416 & 0.572 & -0.707 \\
-0.416 & 0.572 & 0.707 
\end{pmatrix}
.
\ee
The triminimal description of the PMNS 
matrix is then 
\be
U = R_{23}(-\pi/4) \, R_{23} (\epsilon_{23}^\ell) \, R_{13}
(\epsilon_{13}^\ell;\delta^\ell) \, R_{12} (\epsilon_{12}^\ell) \, 
R_{12}(\pi/5)\,.
\ee
We find from this expression that $U_{e3} 
= \sin \epsilon_{13}^\ell \, e^{-i \delta^\ell}$. 
Expanding the PMNS matrix in terms of the small 
parameters gives
\bea \D 
U \simeq U^0 + \epsilon_{12}^\ell
\left(
\bad 
-\frac{\sqrt{5-\sqrt{5}}}{2\sqrt{2}} & 
\frac{1 + \sqrt{5}}{4} & 0 \\ 
-\frac{1 + \sqrt{5}}{4\sqrt{2}} 
& - \frac{\sqrt{5-\sqrt{5}}}{4} & 0 \\ 
-\frac{1 + \sqrt{5}}{4\sqrt{2}} 
& -\frac{\sqrt{5-\sqrt{5}}}{4} & 0 
\ea
\right) + \epsilon_{13}^\ell
\left(
\bad 
0 & 0 & e^{-i \delta^\ell} \\ 
\frac{1 + \sqrt{5}}{4\sqrt{2}} e^{i\delta^\ell} & 
\frac{\sqrt{5-\sqrt{5}}}{4} e^{i\delta^\ell} & 0 \\
-\frac{1 + \sqrt{5}}{4\sqrt{2}} e^{i\delta^\ell} & 
-\frac{\sqrt{5-\sqrt{5}}}{4} e^{i\delta^\ell} & 0 
\ea
\right) \\ \D
+  \epsilon_{23}^\ell
\left(
\bad 
0 & 0 & 0 \\
-\frac{\sqrt{5-\sqrt{5}}}{4} & 
\frac{1 + \sqrt{5}}{4\sqrt{2}} & \sqrt{\frac12} \\
\frac{\sqrt{5-\sqrt{5}}}{4} & 
-\frac{1 + \sqrt{5}}{4\sqrt{2}}  & \sqrt{\frac12}
\ea
\right) .
\eea
Moreover, the other trigonometric functions of the 
mixing angles are 
\be
\sin^2 \theta_{23}^\ell = \frac 12 \, (1 - \sin 2 \epsilon_{23}^\ell) 
\simeq \frac 12 - \epsilon_{23}^\ell
\ee
and 
\be
\ba \D 
\sin^2 \theta_{12}^\ell = \frac{1}{16} \, 
\left(\sqrt{2}\sqrt{5 - \sqrt{5}} \, \cos \epsilon_{12}^\ell 
+ (1 + \sqrt{5}) \sin \epsilon_{12}^\ell \right)^2 \\
\simeq \D 
\frac{5 - \sqrt{5}}{8} + \frac {\sqrt{5 + \sqrt{5}}}{2\sqrt{2}} 
\, \epsilon_{12}^\ell\,.
\ea
\ee
Numerically, 
$\sin^2 \theta_{12}^\ell \simeq 0.345 + 0.951 \, \epsilon_{12}^\ell$, 
so that $\epsilon_{12}^\ell$ is to a good precision the deviation 
of $\sin^2 \theta_{12}^\ell$ from $\frac{5 - \sqrt{5}}{8}$. 
In order to lie in the observed $1\sigma$ range of 
$\sin^2 \theta_{12}^\ell$ given in Eq.~(\ref{eq:datanu}) 
we require 
\be\label{eq:eps12nu}
\epsilon_{12}^\ell = -0.036 \pm 0.020\,.
\ee
With $\epsilon_{12}^\ell = -0.013$ we would obtain 
$\sin^2 \theta_{12}^\ell = \frac 13$. 
The result of the triminimal perturbation of tri-bimaximal mixing, 
i.e., using Eq.~(\ref{eq:PRW}), would be 
$(\epsilon_{12}^\ell)_{\rm TBM} = -0.023 \pm 0.020$. 
This means that the deviations of the tri-bimaximal value 
$\theta_{\rm TBM} \simeq 0.615$ and of $\pi/5$ 
from the best-fit value are of comparable magnitude. 

Comparing Eq.~(\ref{eq:eps12nu}) with the result for the quark sector in 
Eq.~(\ref{eq:eps12q}), we find the remarkable result 
that the best-fit values of the 12-mixing angles are away 
from $\pi/12$ and $\pi/5$ by the same (small) amount 
of roughly\footnote{This number is in magnitude amusingly close 
to the ratio of the 
solar and atmospheric mass-squared differences.} 
$-0.035$. Note however that the error on the leptonic 
parameter $\epsilon_{12}^\ell$ is much larger than that on
$\epsilon_{12}^q$. Nevertheless, this may motivate one to assume 
a unified Ansatz in what regards the 
small parameters of triminimality: 
\be \label{eq:main}
\epsilon_{ij}^q = \epsilon_{ij}^\ell\,.
\ee
This would lead to $\sin^2 \theta_{12}^\ell \simeq 0.313$ and  
$\sin^2 \theta_{23}^\ell \simeq \frac 12 
- |V_{cb}| \simeq 0.458$, which are 
remarkably close to the best-fit values 0.312 and 0.466 quoted in 
Eq.~(\ref{eq:datanu}). The remaining unknown mixing parameter is 
``predicted'' to be $|U_{e3}| = |V_{ub}| 
\simeq 0.00359$ and there is no chance to relate it to the 
(weak) hint for non-zero  $|U_{e3}|$, or to 
measure it in currently planned laboratory experiments. 

Leaving these speculations aside, we have that $\delta^\ell$ is
currently unconstrained, that (at $2\sigma$) 
$\epsilon_{13}^\ell \le 0.036$ and that $\epsilon_{23}^\ell$ 
is to a good precision nothing but the deviation of 
$\sin^2 \theta_{23}^\ell$ from $\frac 12$. To lie in the $1\sigma$ 
($2\sigma$) range one finds 
$(-0.103)-0.039 \le \epsilon_{23}^\ell \le 0.093~(0.136)$. To have 
$\theta_{12}^\ell$ in its allowed $2\sigma$ range we require 
$\epsilon_{12}^\ell = -0.036_{-0.037}^{+0.042}$.  Recall 
that for the quark parameters a hierarchy in the form of 
$(\epsilon_{12}^q)^2 \sim (\epsilon_{23}^q)^2 \sim \epsilon_{13}^q$ 
was implied. In the lepton sector the current lack of comparable 
precision still allows scenarios like 
$|\epsilon_{12}^\ell| \sim |\epsilon_{23}^\ell| 
\sim \epsilon_{13}^\ell$ or $|\epsilon_{23}^\ell| 
\sim (\epsilon_{13}^\ell)^2$. In principle, one, two or even all 
$\epsilon_{ij}^\ell$ could be zero, without being outside the  
allowed $2\sigma$ ranges.

Finally, the invariant 
$J_{\rm CP}^\ell = -{\rm Im} \{U_{e2} \, U_{\mu
2}^\ast \, U_{e 3}^\ast \, U_{\mu3} \}$, which 
governs leptonic CP violation, is found to be 
\bea
J_{\rm CP}^\ell = 
\frac{1}{32 } \cos 2 \epsilon_{23}^\ell \, \sin 2 \epsilon_{13}^\ell
\, \cos \epsilon_{13}^\ell 
    \left( \left(\sqrt{5} - 1\right) 
\sin 2 \epsilon_{12}^\ell + 
\sqrt{2} \, \sqrt{5 + \sqrt{5}} 
\cos 2\epsilon_{12}^\ell \right) \sin \delta^\ell \\
\simeq \D 
\frac 18 
\left(
\frac{\sqrt{5+\sqrt{5}}}{\sqrt{2}} 
+ \epsilon_{12}^\ell \, (\sqrt{5} - 1)
\right) 
\epsilon_{13}^\ell \, \sin \delta^\ell\,.
\eea
In case the relation (\ref{eq:main}) holds we have 
$J_{\rm CP}^\ell \simeq 8.3 \cdot 10^{-4} \,\sin \delta^\ell$. 
It is tempting to assume in addition that $\delta^\ell = \delta^q$,
leading to $J_{\rm CP}^\ell \simeq 7.2 \cdot 10^{-4}$.

One may wonder what kind of Majorana neutrino mass matrix, which is 
defined as 
$m_\nu = U \, {\rm diag}(m_1, m_2, m_3) \, U^T$,
can give rise to $\theta_{23}^\ell = -\pi/4$, $\theta_{13}^\ell = 0$ and 
$\theta_{12}^\ell = \pi/5$. The required form is 
\be
m_\nu = 
\left( 
\bad 
A & B & B \\
\cdot & \frac12 (A + x \, B) + E & \frac12 (A + x \, B) - E \\ 
\cdot & \cdot & \frac12 (A + x \, B) + E
\ea 
\right) ,
\ee
where $x = 2\sqrt{2} \, \sqrt{1 - 2/\sqrt{5}} \simeq 0.919$. 
The neutrino masses are 
$m_1 = A - B \, x \, \sqrt{5}/2$, $m_2 \, e^{2i\alpha}
= A + \sqrt{2} \, \sqrt{1 + 2/\sqrt{5}} \, B$ and 
$m_3 \, e^{2i\beta} = 2 \, E$. 

Simple formulae can also be obtained for the phase-averaged mixing 
probabilities for neutrinos with flavor $\alpha$ to 
end up with flavor $\beta$: 
$\overline{P}_{\alpha \beta} = \sum_i |U_{\alpha i}|^2 \, 
|U_{\beta i}|^2$. This expression is valid when the 
neutrino oscillation length $4 \, E/\Delta m^2$ is much smaller 
than the travelled distance, which is fulfilled, e.g., 
for high energy astrophysical neutrinos. Expanding up to first order 
in the small parameters $\epsilon_{ij}^\ell$ one finds 
\be
\ba \D 
\overline P = \frac{1}{16} \left[ 
\left( 
\bad
11 - \sqrt{5} & \frac12 \, (5 + \sqrt{5}) & \frac12 \, (5 + \sqrt{5}) \\
\cdot & \frac 14 \, (27 - \sqrt{5}) & \frac 14 \, (27 - \sqrt{5}) \\ 
\cdot & \cdot & \frac 14 \, (27 - \sqrt{5}) 
\ea
\right) 
- \epsilon_{12}^\ell \, \sqrt{2} \sqrt{5 - \sqrt{5}} 
\left( 
\bad
4  & -2 & -2 \\
\cdot & 1 & 1 \\ 
\cdot & \cdot & 1
\ea
\right) \right. \\ \D 
\left. 
+ 
\left( 
\cos \delta^\ell \, \epsilon_{13}^\ell \, \sqrt{2} \sqrt{5 - \sqrt{5}}
- \epsilon_{23}^\ell \, (5 + \sqrt{5}) 
\right) 
\left( 
\bad
0 & -1 & 1 \\ 
\cdot & 1 & 0 \\
\cdot & \cdot & -1 
\ea
\right) 
\right] .
\ea 
\ee
As is well-known, the 12-rotation of the lepton sector plays a 
crucial role in \onbb, 
in particular if neutrino masses are inversely hierarchical or
quasi-degenerate. In case of an inverted hierarchy ($m_2 \simeq m_1
\gg m_3$) the effective mass $|(m_\nu)_{ee}|$, on which 
the rate of \onbb~depends quadratically, is 
\be 
|(m_\nu)_{ee}| \simeq \cos^2 \theta_{13} \, \sqrt{\dma} \, 
\sqrt{1 - \sin^2 2 \theta_{12} \, \sin^2 \alpha}\,.
\ee
We have that $\sin^2 2 \theta_{12} \simeq \frac 18 \, (5 + \sqrt{5}) 
+ \epsilon_{12}^\ell \, \sqrt{\frac 12 } \, \sqrt{5 - \sqrt{5}} $, and
the lower limit on $|(m_\nu)_{ee}|$ is 
$\cos^2 \theta_{13} \, \sqrt{\dma} \, \cos 2 \theta_{12}$, where 
$\cos 2 \theta_{12}\simeq  \frac 14 \, (\sqrt{5} - 1) - \epsilon_{12}^\ell
\, \sqrt{\frac 12 } \, \sqrt{5 + \sqrt{5}}$.

As already mentioned, it is not surprising that small numbers 
such as $\theta_{23}^q$ and $\theta_{13}^q$ can be very 
well described by $\pi/n$, where $n \gg 1$. 
It is more the fact that the Cabibbo angle and the 
large mixing angles in the 
lepton sector can also be written as $\pi/n$, which is both 
interesting and for our purposes very helpful. If one insists on the $\pi/n$ 
behavior, then the magnitude of $\theta_{13}^\ell$ 
can only be speculated upon. For instance, 
$\sin^2\theta_{13}^\ell = 0.016$ (0.005) 
corresponds to $\theta_{13}^\ell = \pi/25$ ($\pi/45$).\\

In summary, we have proposed a unified parametrization of the CKM and
PMNS matrices by interpreting the leading mixing angles as being at
zeroth order a fraction of $\pi$ with an integer number. 
While for atmospheric neutrino mixing this is trivial, $\theta_{23}^\ell =
\pi/4$, we have chosen here $\theta_{12}^\ell = \pi/5$, which is consistent 
with the currently allowed $2\sigma$ range, and 
$\theta_{12}^q = \pi/12$. The resulting sines and cosines 
(which correspond to physical quantities) are 
all rather simple, irrational but algebraic numbers. 
We note that $\cos \pi/5 = \varphi/2$, i.e., solar neutrino mixing is 
here connected to the golden ratio $\varphi$. 
The perturbation parameters for $\theta_{12}^\ell$ 
(which are required to reproduce the central values of 
global fits) are of the same order as 
the observed deviations from tri-bimaximal mixing. They are 
small and for the 12-rotations of the same order and sign for 
both quarks and leptons. 


\vspace{0.3cm}
\begin{center}
{\bf Acknowledgments}
\end{center}
I thank Walter Grimus and the Universit\"at Wien, where parts of this 
work were carried out, for kind hospitality 
and am grateful to Claudia Hagedorn as well as 
Lisa Everett for discussions. 
This work was supported by the ERC under the Starting Grant 
MANITOP and by the Deutsche Forschungsgemeinschaft 
in the Transregio 27 ``Neutrinos and beyond -- weakly interacting 
particles in physics, astrophysics and cosmology''.

\end{document}